\documentstyle[preprint,aps]{revtex}
\begin{document}
\draft
\title{
Universal Subgap Optical Conductivity\\ 
in Quasi-One-Dimensional Peierls Systems
}
\author{
Kihong Kim, Ross H. McKenzie, and John W. Wilkins
}
\address{
Department of Physics, The Ohio State
University, Columbus, Ohio  43210\\
}
\maketitle
\begin{abstract}
Quasi-one-dimensional Peierls systems with quantum and thermal 
lattice fluctuations can be modeled by a Dirac-type equation 
with a Gaussian-correlated off-diagonal disorder. 
A powerful new method gives the exact disorder-averaged
Green function used to compute the optical conductivity. 
The strong subgap tail of the conductivity has a universal scaling form.
The frequency and temperature dependence of the calculated spectrum 
agrees with experiments on KCP(Br) 
and {\it trans}-polyacetylene. 
\end{abstract}

\ \


\narrowtext

The interesting new physics exhibited by
quasi-one-dimensional electronic materials continues to
expand as
advances in synthetic chemistry \cite{syn} and semiconductor fabrication
technology \cite{ree} make new
materials available for study.
Many quasi-one-dimensional materials undergo a structural
instability known as the Peierls or charge-density-wave instability
\cite{hkss,gru,car}.
Each chain has
a periodic lattice distortion with twice the Fermi wavevector, $2k_F$,
resulting in a gap $2\Delta$ in the electronic spectrum at the Fermi
surface, suggesting that such a gap should be clearly visible
in the optical absorption spectrum.
However, over the past twenty years, optical absorption
measurements on a wide range of materials \cite{bsz,wei,mcc}
have consistently shown that there is a broad tail below the gap.
Moreover, as the temperature increases the spectrum broadens
considerably.
There is
is no accepted quantitative theory of the shape of the 
spectrum or its temperature dependence.
While considerable effort has been made to understand
excitation energies of solitons and polarons \cite{hkss}
in these compounds, there is also no accepted theory for 
their absorption spectrum.

The zero-point and thermal lattice motions
have a  significant effect on the electronic properties of
the Peierls state and are, we believe, key to understanding the subgap absorption
\cite{mw}.  A recent luminescence and Raman study of a
metal halogen mixed-valence (MX chain)
compound produced results consistent with this view \cite{lon}.
If the relevant phonon frequency at $q=2k_F$, $\omega_{2k_F}$,  
is much smaller than the optical frequency, as is the
case in the optical absorption in most Peierls materials,  
the quantum and thermal lattice fluctuations can be modeled \cite{mw,braz}
by a static, Gaussian-random, backscattering potential,  
$\xi(x)$, with zero mean.
Further, if the phonon dispersion near $q=2k_F$ is ignored, $\xi(x)$ is
fully characterized by the disorder-averaged correlator  
$\langle\xi(x)\xi^*(y)\rangle=\gamma\delta(x-y)$.
In general, the dimensionless disorder parameter 
$\eta\equiv\gamma/\hbar v_F\Delta$, where $v_F$ is the Fermi velocity,
has contributions from extrinsic disorder such as due to impurities,
$\eta_e$, as well as from the intrinsic disorder due to
lattice fluctuations, $\eta_i$.
The latter
is related to the dimensionless electron-phonon coupling constant
$\lambda$, the phonon frequency at $2k_F$, and temperature \cite{mw}:
\begin{equation}
\eta(T)=\eta_e+\eta_i(T)=\eta_e+\lambda\frac{\pi\hbar\omega_{2k_F}}{2\Delta}
\coth\left( \frac{\hbar\omega_{2k_F}}{2k_BT}\right).
\label{eta}
\end{equation}

In this Letter, we develop a novel and powerful method for computing the 
electronic properties of one-dimensional Peierls semiconductors
with static disorder. We use this method to calculate the real part of 
the optical conductivity.
We find (1) a strong
subgap tail in the optical absorption;
(2) the remarkable fact
that the \underline{sub}gap conductivity, when properly scaled,
follows a universal scaling curve independent of the disorder
parameter $\eta$;
(3) good agreement between the computed conductivity
and the experimental conductivity for KCP(Br) \cite{bsz}
and {\it trans}-polyacetylene \cite{wei};
(4) a universal scaling curve for the experimental conductivity 
for KCP(Br) at different 
temperatures  
with $\eta$ having the temperature dependence given by Eq.~(\ref{eta}).
Figure 1 contrasts the relatively good agreement with experiment achieved
by our calculation against that for a Lorentzian convolution of the
rigid-lattice conductivity \cite{lc}.

We consider the standard continuum model \cite{braz,maki} of noninteracting electrons in one 
dimension with a Peierls ground state, which, without the random term,
has been studied extensively (see p. 795 in \cite{hkss}) 
to understand the soliton and polaron
excitations in conducting polymers. Our system is described by
a Dirac-type equation \cite{ov} for the wave functions
$\psi_1(x)$ and $\psi_2(x)$ for electrons moving 
with $v_F$ to the right and to
the left, respectively, in the interval $-L\leq x\leq L$ 
with a complex, Gaussian-random, backscattering potential $\xi(x)$:
\begin{eqnarray}
& &\left( \begin{array}{cc} -i\hbar v_F\displaystyle{\frac{\partial}
{\partial x}}& \ \Delta+\xi(x)\\
\Delta+\xi^*(x)& \ i\hbar v_F\displaystyle{\frac{\partial}{\partial x}}
\end{array} \right)
\left( \begin{array}{c} \psi_1(x)\\ \psi_2(x)\end{array}\right)
=E\left(\begin{array}{c} \psi_1(x)\\ \psi_2(x)\end{array}\right),\nonumber \\
& &\langle\xi(x)\rangle=\langle\xi(x)\xi(y)\rangle=0,
~\langle\xi(x)\xi^*(y)\rangle=\gamma
\delta(x-y).
\label{main}
\end{eqnarray}
The above equations
correspond to the case where the Fermi wavelength of the electron is
incommensurate with the lattice period \cite{model}.
Previous work has succeeded in computing the density of states
and localization length for this model \cite{abr}.
Here we report the first exact calculation of the Green function \cite{dos}.

We use $G^+~(G^-)$, the retarded (advanced) $2\times 2$ matrix 
Green function 
to compute the real part of the 
frequency-dependent conductivity for $\hbar\omega\gg k_BT$ \cite{ov,rs}:
\begin{equation}
\sigma(\omega)=\frac{e^2}{A\hbar}\frac{2}{\pi\hbar\omega}
\int_{-\hbar\omega}^0 dE\int_y^\infty dx~ 
{\rm Re}\left[j^{+-}(E+\hbar\omega,E|
x-y)
-j^{++}(E+\hbar\omega,E|x-y)\right],
\end{equation}
where $A$ is the cross-sectional area of a chain and
\begin{equation}
j^{\pm\pm}(E^\prime,E|x-y)=(\hbar v_F)^2\big\langle 
{\rm Tr}\left[\sigma_3 G^\pm(y,x|E^\prime)
\sigma_3 G^\pm(x,y|E)
\right]\big\rangle .
\end{equation}
$\sigma_3$ is a Pauli matrix.
Here we report the computation of 
the conductivity using the approximation 
$\langle GG\rangle\approx \langle G\rangle\langle G\rangle$,
which we expect to be valid for $\hbar\omega \gg \gamma /\hbar v_F$,
i.e., for optical but not dc conductivities.  
However, our method can be used to compute $\langle GG\rangle$ 
directly.

{\it Exact calculation of the Green function.}~~~Two linearly independent wave 
functions $\psi$ and $\tilde\psi$ satisfy the boundary conditions:
$\psi(-L)=1$ and $\tilde\psi(L)=1$. These $\psi$ and $\tilde\psi$ 
will be statistically independent in the limit $L\rightarrow
\infty$ \cite{ov}. For $x>y$,
\begin{equation}
G(x,y|E)=\frac{i}{\hbar v_F(\psi_1\tilde\psi_2-\psi_2\tilde\psi_1)}
\left(
\begin{array}{cc} \tilde\psi_1(x)\psi_2(y)&\tilde\psi_1(x)\psi_1(y)\\
\tilde\psi_2(x)\psi_2(y)&\tilde\psi_2(x)\psi_1(y) \end{array}\right),
\end{equation}
where ($\psi_1\tilde\psi_2-\psi_2\tilde\psi_1$) is a constant
independent of $x$.
We obtain $G^+~(G^-)$ by solving for wave functions with Im$E$ a small
positive (negative) number. Then the denominator can be Taylor-expanded 
and 
\begin{equation}
G^+(x,y|E)=-\frac{i}{\hbar v_F}\sum_{n=0}^\infty\left(
\begin{array}{cc}
\hat C_n(x,y)\hat y_n(x)& \hat D_n(x,y)\hat y_n(x)\\
\hat C_n(x,y)\hat y_{n+1}(x)& \hat D_n(x,y)\hat y_{n+1}(x)
\end{array}\right),
\label{rgf}
\end{equation}
where
\begin{equation}
\hat y_n(x)=\left[\frac{\tilde\psi_2(x)}{\tilde\psi_1(x)}
\right]^n,~\hat C_n(x,y)=\left[\frac{\psi_1(x)}{\psi_2(x)}
\right]^n\frac{\psi_2(y)}{\psi_2(x)},~\hat D_n(x,y)=
\left[\frac{\psi_1(x)}{\psi_2(x)}\right]^n\frac{\psi_1(y)}
{\psi_2(x)}.
\label{ic}
\end{equation}

Next, we take the $L\rightarrow\infty$ limit and utilize the statistical 
independence of $\psi$ and $\tilde\psi$ to factor the average of
the Green function into products of averages of 
$\hat C$ ($\hat D$) and $\hat y$.
The quantities $C_n\equiv\langle\hat C_n(x,y)\rangle$ 
and $D_n\equiv
\langle\hat D_n(x,y)\rangle$ are functions of $(x-y)$ and $y_n\equiv
\langle\hat y_n(x)\rangle$  is independent of $x$.
Thus the averaged Eq.~(\ref{rgf})
can be written in terms of $y_n$, $C_n(x-y)$ and $D_n(x-y)$.
The following equations for $y_n$ and $C_n$ can be derived 
from Eq.~(\ref{main}) using standard Fokker-Planck
methods \cite{halp},
subject to the conditions $y_0=1$ and $C_n(0)=y_n$,
which follow from Eq.~(\ref{ic}) for $y_0$, and from Eq.~(\ref{ic}) and the equality
$\langle(\tilde\psi_2/\tilde\psi_1)^n\rangle=\langle(\psi_1/\psi_2)^n
\rangle$ \cite{ov} for $C_n(0)$:
\begin{equation}
\left(2\frac{E}{\Delta}+i\eta n\right)y_n-y_{n+1}-y_{n-1}=0,
\label{y}
\end{equation}
\begin{equation}
\frac{\hbar v_F}{\Delta}\frac{dC_n}{d(x-y)}=i(2n+1)\frac{E}{\Delta}C_n
-i(n+1)C_{n+1}-inC_{n-1}-\eta\left(n^2+n+\frac{1}{2}\right)C_n.
\label{c}
\end{equation}
$D_n$ satisfies the same equation as $C_n$ with the initial condition
$D_n(0)=y_{n+1}$.

{\it The key to our method is the fact that $y_n$, $C_n$ and $D_n$
decay very rapidly and exponentially as a function of $n$.}
The decay rate is determined by the dimensionless 
disorder parameter $\eta=\gamma/\hbar v_F\Delta$:
the larger $\eta$, the more rapid the decay.
To solve Eqs.~(\ref{y}) and (\ref{c}) numerically,
we truncate the number of equations by setting $y_n=0$
for $n>N$. For example, at $\eta=0.1$, the Green function is 
converged to 1 ppm for $N=30$. The method can be generalized to
the calculation of $\langle GG\rangle$.
The number of resulting equations scales
as $N^2$.

We calculate the conductivity using the exact Green function
obtained by solving Eqs.~(\ref{y}) and (\ref{c}) and the approximation
$\langle GG\rangle\approx\langle G\rangle\langle G\rangle$. 
In Fig.~2(a), we show the result for several different
$\eta$ values. The $\eta=0$ curve shows
no absorption below the energy gap and a divergent behavior
just above the gap.
For finite $\eta$ values, the singularity is absent, and the spectrum
broadens rapidly as $\eta$ increases.
In the inset of Fig.~2(b), we plot the dependence on $\eta$ of three
quantities characterizing the shape of the conductivity curve: 
$\Gamma$, which is the half-width for the low frequency side
of the peak, the inverse of the conductivity at the peak, and
$\delta\omega$, the shift of the peak frequency from $2\Delta$.
Quite remarkably, when we scale the conductivity by the peak value
and the frequency by $\Gamma$, we find that the scaled curves for 
all $\eta$ values have {\it a universal form independent of $\eta$}
below the peak frequency \cite{urb,uni}.

Experimental data on the platinum, linear-chain compound
K$_2$Pt(CN)$_4$Br$_{0.3}$$\cdot$3H$_2$O or KCP(Br) \cite{bsz}
show a similar scaling behavior.
In Fig.~3, we plot the data for five temperatures. The scaling
curve below the conductivity peak has approximately the same form as
the theoretical curve in Fig.~2(b). We deduce the effective disorder parameter
for each temperature by choosing $\eta$ such that the computed
$\Gamma(\eta)/\omega_{\rm peak}(\eta)$ \cite{univ} 
is equal to $\Gamma(T)/\omega_{\rm peak}
(T)$. This criterion was used to choose the value $\eta=0.43$ in Fig.~1.
In the inset of Fig.~3 we plot the $\eta$ values thus determined versus temperature.
The temperature dependence of $\eta$ is compared 
with the postulated form (\ref{eta}): an extrinsic temperature-independent
$\eta_e$ plus an intrinsic temperature-dependent $\eta_i$.
The fit value of extrinsic disorder is large. 
This is not surprising since x-ray and neutron scattering studies have shown that the
bromine ions randomly occupy two different sites \cite{wil}.
This disorder could cause significant 
scattering of the electrons along the platinum chains \cite{wag}.

We have analyzed the optical conductivity data for {\it trans}-polyacetylene
(CH)$_x$ at room temperature \cite{wei}. Since the phonon frequency
at $2k_F$ for this material is at a much larger energy scale than 
the room temperature, the optical conductivity does not depend 
sensitively on temperature. Consequently, unlike in KCP(Br), we
cannot determine the separate contributions from extrinsic and intrinsic
disorder.
In Fig.~3, we show that the scaled (CH)$_x$ \underline{sub}gap
conductivity curve agrees well
with those of KCP(Br). 

The peak of the conductivity is the product of $\sigma_0\equiv e^2v_F/A\Delta$
and the dimensionless ratio $\sigma_{\rm peak}(\eta)/\sigma_0$
determined from fitting the experimental curves. 
Accordingly the peak conductivity yields $\sigma_0$ and hence $v_F$
since the gap $2\Delta$ and cross-sectional area $A$ are known. Table I 
shows that the $v_F$'s obtained this way are smaller than
estimates based on band structure calculations.
The values of $\eta_i(T=0)$ for KCP(Br) and $\eta(T=0)$ for
(CH)$_x$ obtained from the fitting are compared with
estimates from the lattice zero-point motion.

In conclusion, we have developed a new method for computing the electronic 
properties of quasi-one-dimensional Peierls systems including
disorder and lattice fluctuation effects and used it to
obtain the absorptive conductivity at all temperatures. Our results 
compare well with data on KCP(Br) and {\it trans}-polyacetylene. 
The \underline{sub}gap conductivity curve has a universal scaling form. 

We thank Leo Degiorgi for helpful discussions and showing us his results 
before publication.
This research has been supported in part by the US DOE-Basic Energy
Sciences, Division of Materials Sciences and the OSU Center for 
Materials Research.

\begin{figure}
\caption{Comparison of the real part of the frequency-dependent
conductivity, $\sigma(\omega)$, obtained from
our calculations for a disorder
parameter $\eta=0.43$ with experiment on KCP(Br) at $T=40$ K 
\protect\cite{bsz} and the conductivity obtained from a Lorentzian
convolution of the rigid-lattice conductivity 
for a dimensionless damping parameter $\zeta=0.42$ \protect\cite{lc}.
The frequency and the conductivity are scaled by the frequency at 
the peak, $\omega_{\rm peak}$, and by $\sigma_{\rm peak}=\sigma(
\omega_{\rm peak})$, respectively.}
\end{figure}

\begin{figure}
\caption{Universal form of the computed subgap optical conductivity. 
(a) Conductivity spectrum for different values
of the disorder parameter $\eta$ (Eq.~(\protect\ref{eta})). 
The spectrum broadens as $\eta$ increases.
The conductivity scale is set by $\sigma_0=e^2 v_F/A \Delta$,
where $A$ is the cross-sectional area of a chain.
(b) Scaling plot of $\sigma(\omega)$,
where the frequency is scaled by $\Gamma(\eta)$, 
the half-width for the low frequency side of the peak.
Note we claim only that the \underline{sub}gap ($\omega<\omega_{\rm peak}$)
conductivity scales.
Inset: Dependence on the disorder parameter $\eta$ 
of $\sigma_{\rm peak}$, $\Gamma$, and $\delta\omega
(\eta)\equiv 2\Delta-\omega_{\rm peak}(\eta)$
.}
\end{figure}

\begin{figure}
\caption{Scaling plot of experimental data for KCP(Br) \protect\cite{bsz}
and {\it trans}-(CH)$_x$ \protect\cite{wei} at different 
temperatures. 
For each temperature, the corresponding disorder
parameter value is obtained by choosing $\eta$ such that 
the theoretical $\Gamma(\eta)/\omega_{\rm peak}(\eta)$ value is
equal to $\Gamma(T)/\omega_{\rm peak}(T)$.
Inset: Comparison of the temperature dependence of $\eta$ for KCP(Br) 
with Eq.~(\protect\ref{eta}).
We fit the data with a curve in which the first number (0.3) 
characterizes extrinsic disorder effects and the second number
($\eta_i(0)=0.07$) depends sensitively on the value of $\omega_{2k_F}$
(twice the third number in K) which we took from 
experiment \protect\cite{car}.
The possibility of the scaling curve extending to 
temperatures above the three-dimensional ordering transition at
$T_P=$ 120 K \protect\cite{car} is based on the 
persistence of a `pseudogap' above $T_P$ \protect\cite{rice}.
}
\end{figure}

\begin{table}
\squeezetable
\caption{Fermi velocity $v_F$ and 
disorder parameter $\eta$ determined from fits of
experimental optical conductivity to our calculations.
The peak height determines $v_F$ which is compared to the results
of band structure calculations
\protect\cite{hkss,car}.
The $\eta$ values obtained from the fits are compared
to the intrinsic $\eta_i(T=0)=(\delta u/ u_o)^2 (\Delta a / \hbar v_F)$,
calculated using the band structure $v_F$,
and the zero-point motion $\delta u$ and lattice
distortion $u_o$ in Table I of 
Ref.~\protect\cite{mw}.
The electron-phonon coupling $\lambda$ is calculated from the
estimated value of $\eta_i$ and Eq.~(1). 
}
\begin{tabular}{lccccc}
  &\multicolumn{2}{c}{$v_F \  (10^7 {\rm cm/sec})$} &\multicolumn{2}{c}
  {$\eta$} &\\
   & Fit & Band structure  &Fit & Estimate of $\eta_i$ &$\lambda$\\
\tableline
KCP(Br)& 5.2 & 11.0 & $\eta_e$=0.3, $\eta_i(0)$=0.07 & 0.07 &0.96\\
(CH)$_x$ & 7.1 & 9.3 & $\eta_e$+$\eta_i(0)$=0.15  & 0.18 &0.48\\
\end{tabular}
\label{table1}
\end{table}


\begin{references}

\bibitem{syn} Proceedings of the International Conference on
Science and Technology of Synthetic Metals,
G\"oteberg, Sweden, August 12-18, 1992, Synth. Met.
{\bf 55}--{\bf 57} (1993).

\bibitem{ree} For a discussion of quantum wires see, e.g.,
C. Weisbuch and B. Vinter,
{\it Quantum Semiconductor Structures} 
(Academic, Boston, 1991), Ch.~6.

\bibitem{hkss} A. J. Heeger, S. Kivelson, J. R. Schrieffer, and
W.-P. Su, Rev. Mod. Phys. {\bf 60}, 781 (1988).
For (CH)$_x$, $v_F=2 t_o a/\hbar$ where $4t_o=$ 10 eV is 
the tight-binding bandwidth and $a=1.24~\AA$ is the intrachain atomic separation.
The cross-sectional area per chain $A=1.55 \times 10^{-15}~{\rm cm}^2$.

\bibitem{gru} G. Gr\"uner,
Rev. Mod. Phys. {\bf 60}, 1129 (1988).

\bibitem{car} K. Carneiro, in {\it Electronic Properties of Inorganic
Quasi-One-Dimensional Compounds, Part II}, edited by P. Monceau (Reidel,
Dordrecht, 1985), p.~1.
KCP(Br) has a free-electron-like band, $v_F= \hbar k_F /m $.
$k_F= 0.94~\AA^{-1}$.
$A=4.90 \times 10^{-15}~{\rm cm}^2$.

\bibitem{bsz} P. Br\"uesch, S. Str\"assler, and H. R. Zeller, Phys. Rev.
B {\bf 12}, 219 (1975) [KCP(Br)].

\bibitem{wei} G. Leising, Phys. Rev. B {\bf 38}, 10313 (1988)
[(CH)$_x$].

\bibitem{mcc}
R. P. McCall {\it et al.},
Phys. Rev. B {\bf 39}, 7760 (1989) [TCNQ compounds];
D. Berner {\it et al.},
J. Phys. (France) IV C2 {\bf 3}, 255 (1993) [(TaSe$_4$)$_2$I];
L. Degiorgi, J. Phys. (France) IV C2 {\bf 3}, 103 (1993) [K$_{0.3}$MoO$_3$].

\bibitem{mw} R. H. McKenzie and J. W. Wilkins,
Phys. Rev. Lett. {\bf 69}, 1085 (1992); R. H. McKenzie and J. W. Wilkins,
Synth. Met. {\bf 55}--{\bf 57}, 4296 (1993).

\bibitem{lon} F. H. Long, S. P. Love, B. I. Swanson, and  R. H. McKenzie,
Phys. Rev. Lett. {\bf 71}, 762 (1993).

\bibitem{braz} S. A. Brazovskii and I. E. Dzyaloshinskii, Zh. Eksp. Teor. 
Fiz. {\bf 71}, 2338 (1976) [Sov. Phys. JETP {\bf 44}, 1233 (1976)].

\bibitem{lc} If we convolute the rigid-lattice conductivity with a 
Lorentzian, we obtain
\begin{displaymath}
\sigma(\omega)=\frac{2}{\pi}\frac{\Delta}{\omega}{\rm Im}\int_0^\infty
dx~\frac{1}{\sqrt{1+x^2}[1+x^2-(\omega/\Delta+i\zeta)^2/4]},
\end{displaymath}
where $\zeta\Delta$ corresponds to an energy-independent damping 
of the transition. 
Z.-b. Su and L. Yu [Commun. Theor. Phys. (Beijing) {\bf 2}, 1341 (1983)] previously
pointed out that this form does not agree quantitatively with the observed
subgap absorption.

\bibitem{maki} H. Takayama, Y. R. Lin-Liu, and K. Maki, 
Phys. Rev. B {\bf 21}, 2388 (1980).

\bibitem{ov} A. A. Ovchinnikov and N. S. \'Erikhman, 
Zh. Eksp. Teor. Fiz. {\bf 78}, 1448 (1980) 
[Sov. Phys. JETP {\bf 51}, 728 (1980)].

\bibitem{model} In particular, 
the condition $\langle\xi(x)\xi(y)\rangle=0$ 
implies that there is no umklapp scattering. 
A non-zero value occurs for a {\it real} random backscattering 
potential which corresponds to the 
half-filled band (commensurate) case, 
an example of which is polyacetylene (CH)$_x$.

\bibitem{abr} 
H. J. Fischbeck and R. Hayn, Phys. Stat. Sol. (b) {\bf 158},
565 (1990), and references therein.

\bibitem{dos} Our method can be further generalized to the case with
forward, backward, and umklapp scattering processes. In fact, we
can reproduce the results of previous density of states
calculations \cite{abr} with great efficiency.

\bibitem{rs} J. Rammer and H. Smith, Rev. Mod. Phys. {\bf 58}, 323 (1986).

\bibitem{halp} 
I. M. Lifshits, S. A. Gredeskul, and L. A. Pastur, {\it Introduction to
the Theory of Disordered Systems} (Wiley, New York, 1988), p.~146.

\bibitem{urb} 
This universal behavior is quite distinct from that obeyed by the 
Urbach tails seen in three-dimensional crystalline and amorphous
semiconductors.

\bibitem{uni} The scaling curve in Fig.~2(b) below the absorption peak 
is fitted with the function
$\sigma(\omega)/\sigma(\omega_{\rm peak})=\exp[-0.49|(\omega-\omega_{\rm peak})/
\Gamma|^2-0.20|(\omega-\omega_{\rm peak})/\Gamma|^3]$. 

\bibitem{univ} The dependence of $\Gamma/\omega_{\rm peak}$ 
on the disorder parameter $\eta$
can be fitted to 
$\Gamma(\eta)/\omega_{\rm peak}(\eta)=\eta^{0.62}(0.414 + 0.077 \eta)$.

\bibitem{wil} J. M. Williams {\it et al.}, 
Phys. Rev. Lett. {\bf 33}, 1079 (1974).

\bibitem{wag} H. Wagner et al., Solid State Commun. {\bf 13},
659 (1973).

\bibitem{rice} 
P. A. Lee, T. M. Rice, and P. W. Anderson,
Phys. Rev. Lett. {\bf 31}, 462 (1973); M. J. Rice and S. Str\"assler,
Solid State Commun. {\bf 13}, 1389 (1973).

\end{references}
\end{document}